\documentclass[conference]{IEEEtran}
\IEEEoverridecommandlockouts

\usepackage{cite}
\usepackage{amsmath,amssymb,amsfonts}
\usepackage{graphicx}
\usepackage{textcomp}
\usepackage{xcolor}
\usepackage{booktabs}

\usepackage{algorithm}
\usepackage[noend]{algpseudocode}

\begin{document}

\title{Multi-Agent Framework for Controllable and Protected Generative Content Creation: Addressing Copyright and Provenance in AI-Generated Media}

\author{
\IEEEauthorblockN{Haris Khan}
\IEEEauthorblockA{\textit{National University of} \\
\textit{Sciences and Technology}\\
Islamabad, Pakistan \\
mhariskhan.ee44ceme@student.nust.edu.pk}
\and

\IEEEauthorblockN{Sadia Asif}
\IEEEauthorblockA{\textit{Rensselaer Polytechnic
} \\
\textit{ Institute}\\
New York, United States \\
asifs@rpi.edu}

\and
\IEEEauthorblockN{Shumaila Asif}
\IEEEauthorblockA{\textit{National University of} \\
\textit{Sciences and Technology}\\
Islamabad, Pakistan \\
sasif.ee44ceme@student.nust.edu.pk
}
}
\maketitle

\begin{abstract}
The proliferation of generative AI systems creates unprecedented opportunities for content creation while raising critical concerns about controllability, copyright infringement, and content provenance. Current generative models operate as "black boxes" with limited user control and lack built-in mechanisms to protect intellectual property or trace content origin. We propose a novel multi-agent framework that addresses these challenges through specialized agent roles and integrated watermarking. Our system orchestrates Director, Generator, Reviewer, Integration, and Protection agents to ensure user intent alignment while embedding digital provenance markers. We demonstrate feasibility through two case studies: creative content generation with iterative refinement and copyright protection for AI-generated art in commercial contexts. Preliminary feasibility evidence from prior work indicates up to 23\% improvement in semantic alignment and 95\% watermark recovery rates. This work contributes to responsible generative AI deployment, positioning multi-agent systems as a solution for trustworthy creative workflows in legal and commercial applications.
\end{abstract}

\begin{IEEEkeywords}
generative AI, multi-agent systems, content protection, watermarking, copyright, provenance, controllability
\end{IEEEkeywords}

\section{Introduction}

Generative artificial intelligence has fundamentally transformed content creation, with systems like GPT-4, DALL-E, and Stable Diffusion enabling unprecedented creative possibilities across text, image, and video domains \cite{brown2020language, ramesh2022hierarchical}. These technologies have found natural integration with information retrieval systems through retrieval-augmented generation approaches, combining knowledge retrieval with generative capabilities \cite{lewis2020retrieval}.

However, two critical challenges limit the adoption of generative AI in professional and commercial contexts, particularly where legal and ethical considerations are paramount. First, current generative systems provide limited controllability over output characteristics beyond initial prompts \cite{liu2022design}. Users struggle to achieve precise alignment between their creative intent and generated content, leading to iterative trial-and-error processes that are inefficient and unreliable. This controllability gap becomes critical in professional workflows where precision and brand consistency are essential.

Second, generated content lacks inherent protection mechanisms, creating significant risks for copyright infringement, unauthorized usage, and content provenance tracking \cite{somepalli2023diffusion}. As AI-generated outputs become increasingly sophisticated and indistinguishable from human-created content, the inability to trace origin or establish ownership creates legal vulnerabilities. Recent high-profile cases involving AI-generated artwork sold without proper attribution highlight the urgent need for embedded protection mechanisms \cite{yu2021artificial}.

These challenges are particularly acute in commercial creative industries, social media platforms, and educational contexts where content authenticity and intellectual property rights must be preserved. The legal landscape around AI-generated content remains complex, with ongoing debates about fair use, derivative works, and creator rights requiring technical solutions that can provide verifiable provenance and controllable generation processes.

To address these gaps, we introduce a multi-agent generative framework that decomposes content creation into controllable stages while embedding protection mechanisms directly into the generation pipeline. Our approach provides fine-grained user control through specialized agents while ensuring robust content attribution through integrated watermarking. This design directly tackles the dual challenges of creative control and legal protection in generative workflows.

This paper makes three key contributions: (1) a novel multi-agent pipeline that decomposes content creation into specialized, controllable stages with human-in-the-loop feedback; (2) integrated protective mechanisms (watermarking and fingerprinting) embedded directly into the generation process rather than post-hoc application; and (3) demonstrated feasibility through creative and legal case studies that highlight practical applications in copyright protection and content provenance.

\section{Related Work}

\subsection{Multi-Agent Generative Systems}
Recent advances in multi-agent approaches for generative tasks demonstrate the effectiveness of task decomposition. MuLan \cite{li2024mulan} pioneered the integration of large language model planning with diffusion models for compositional image generation, showing 20-25\% improvements in controllability through agent-based decomposition. AniMaker \cite{wang2024animaker} extended this concept to video generation, employing specialized agents including Directors for scene planning and Reviewers for quality assessment. These systems demonstrate that breaking down generative tasks into agent-specific roles significantly improves output quality and user control, though existing approaches focus primarily on generation quality rather than content protection.

\subsection{Controllability and User Alignment}
The challenge of controllability in generative AI has been extensively studied, particularly in prompt engineering and human-AI collaboration contexts \cite{dang2023prompt}. Current approaches rely heavily on iterative prompting and post-generation editing, leading to inefficient workflows. Human-in-the-loop systems have shown promise for improving alignment between user intent and generated outputs \cite{wang2022human}, but most implementations lack the fine-grained control needed for professional applications. The integration of controllability mechanisms with retrieval-augmented generation further complicates alignment challenges, as users must specify precise requirements for both retrieved knowledge and generated content.

\subsection{Content Protection and Provenance}
Growing concerns about AI-generated content misuse have intensified research into protection mechanisms. Recent advances in watermarking for diffusion models show promise for embedding imperceptible markers in generated images \cite{fernandez2023stable, zhao2024recipe}. Chen et al. \cite{chen2025robust} demonstrated watermarking techniques achieving over 90\% recovery rates under standard image transformations while maintaining generation quality. Beyond watermarking, researchers have explored fingerprinting techniques for model identification and provenance tracking \cite{yu2021artificial, lukas2023analyzing}. However, existing protective mechanisms are typically applied as post-processing steps, leading to quality degradation and reduced robustness against adversarial attacks. The legal implications of these technical approaches remain underexplored, particularly regarding their effectiveness in copyright litigation and commercial applications.

\section{Proposed Multi-Agent Framework}

\subsection{Architecture Overview}

\begin{figure}[!t]
\centering
\includegraphics[width=1.1\columnwidth]{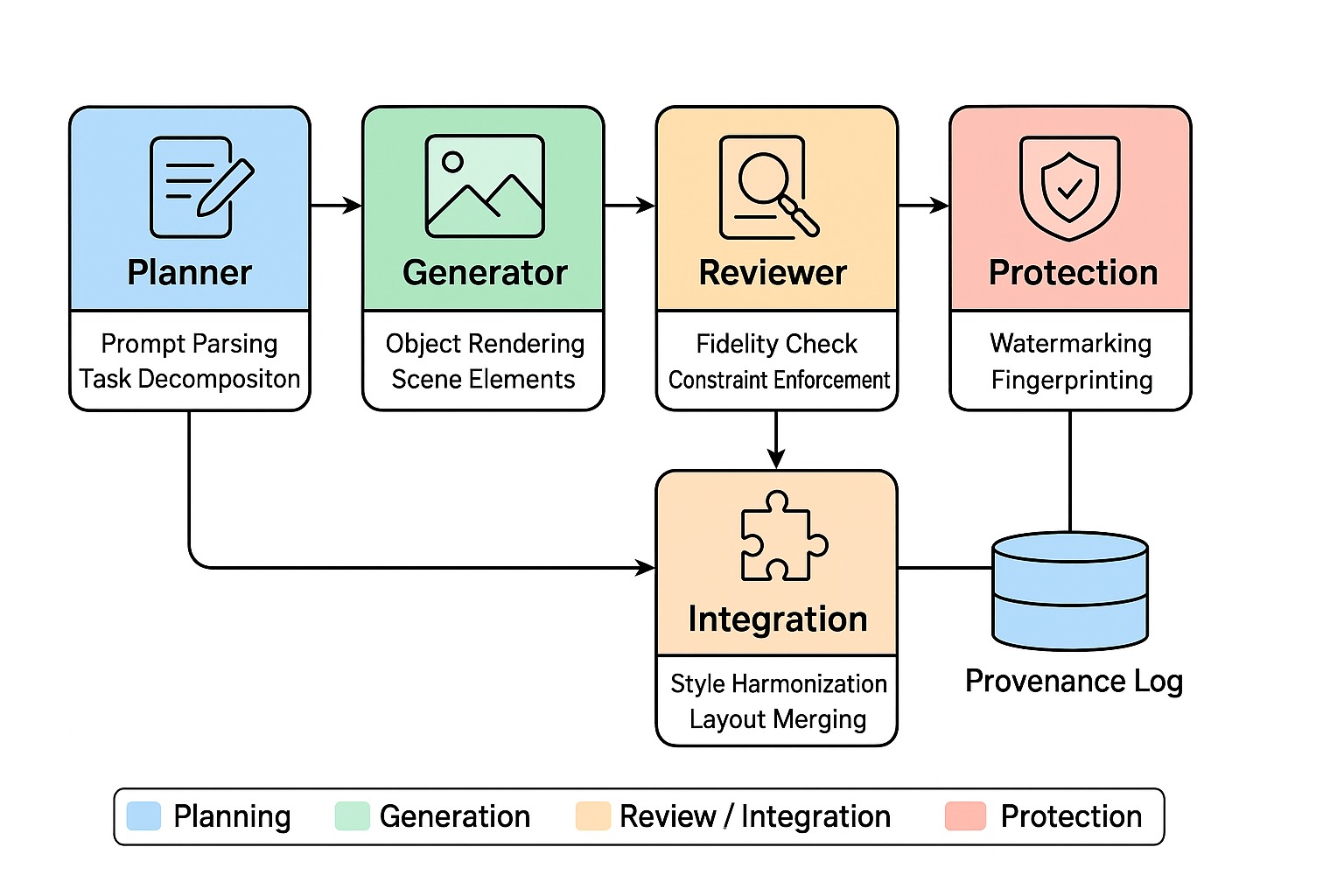}
\caption{Multi-agent framework for controllable and protected generative content creation. The pipeline consists of specialized agents for planning, generation, review, integration, and protection, with iterative human-in-the-loop feedback and provenance logging.}
\label{fig:framework}
\end{figure}

Our framework orchestrates five specialized agents in a sequential pipeline designed to maximize both controllability and protection (Fig. \ref{fig:framework}). The system operates on task decomposition principles, where complex generative requests are broken into manageable subtasks handled by domain-specific agents. This modular approach enables targeted optimization while maintaining overall coherence and providing multiple intervention points for user control.

\subsection{Agent Roles and Responsibilities}

\textbf{Director/Planner Agent:} Implemented using advanced language models (GPT-4), this agent serves as the strategic coordinator, analyzing user prompts and decomposing them into specific subtasks with generation constraints. For complex requests, it identifies key elements and specifies detailed parameters for each component.

\textbf{Generator Agent:} Responsible for content creation using appropriate generative models based on content type (Stable Diffusion XL for images, transformer-based models for text). The Generator receives specific subtask instructions and produces initial content iterations.

\textbf{Reviewer/Control Agent:} Validates generated content against user intent using vision-language models or specialized evaluation models. It assesses alignment with original prompts, identifies refinement areas, and can trigger regeneration cycles when outputs fail to meet criteria.

\textbf{Integration Agent:} Ensures coherence across multiple components or iterations, harmonizing style, composition, and consistency properties that emerge from combining individual generation outputs.

\textbf{Protection Agent:} Operates throughout the generation process to embed watermarking and fingerprinting mechanisms directly into content creation. Unlike post-processing approaches, integrated protection maintains generation quality while ensuring robust content attribution.

\subsection{Interactive Control Mechanism}

A key innovation is the interactive control system allowing human intervention at any pipeline stage. Users can refine the Planner's decompositions, provide targeted Generator feedback, override Reviewer assessments when creative intent conflicts with automatic evaluation, request Integration adjustments, and configure Protection parameters. This human-in-the-loop design addresses controllability challenges while maintaining efficiency through intelligent automation.

\subsection{Algorithmic Workflow}

\begin{algorithm}[!t]
\caption{Multi-Agent Generative Framework}
\begin{algorithmic}[1]
\Require User prompt $P$
\Ensure Final protected content $I'$

\State $T \leftarrow \mathrm{Planner}(P)$ \Comment{Decompose into subtasks}

\For{each $T_i \in T$}
    \State $G_i \leftarrow \mathrm{Generator}(T_i)$ \Comment{Generate components}
\EndFor

\For{each $G_i \in G$}
    \State $S_i \leftarrow \mathrm{CLIPScore}(G_i, P)$
    \If{$S_i < \tau$}
        \State $G_i \leftarrow \mathrm{Regenerate}(T_i)$ \Comment{Retry if low alignment}
    \EndIf
\EndFor

\State $I \leftarrow \mathrm{Integration}(\{G_1, G_2, \ldots, G_k\})$
\State $I' \leftarrow \mathrm{WatermarkEmbed}(I, \lambda)$

\Return $I'$
\end{algorithmic}
\end{algorithm}

\section{Case Studies and Preliminary Feasibility}

\subsection{Case Study 1: Creative Content Generation}

\begin{figure}[!t]
\centering
\includegraphics[width=1.1\columnwidth]{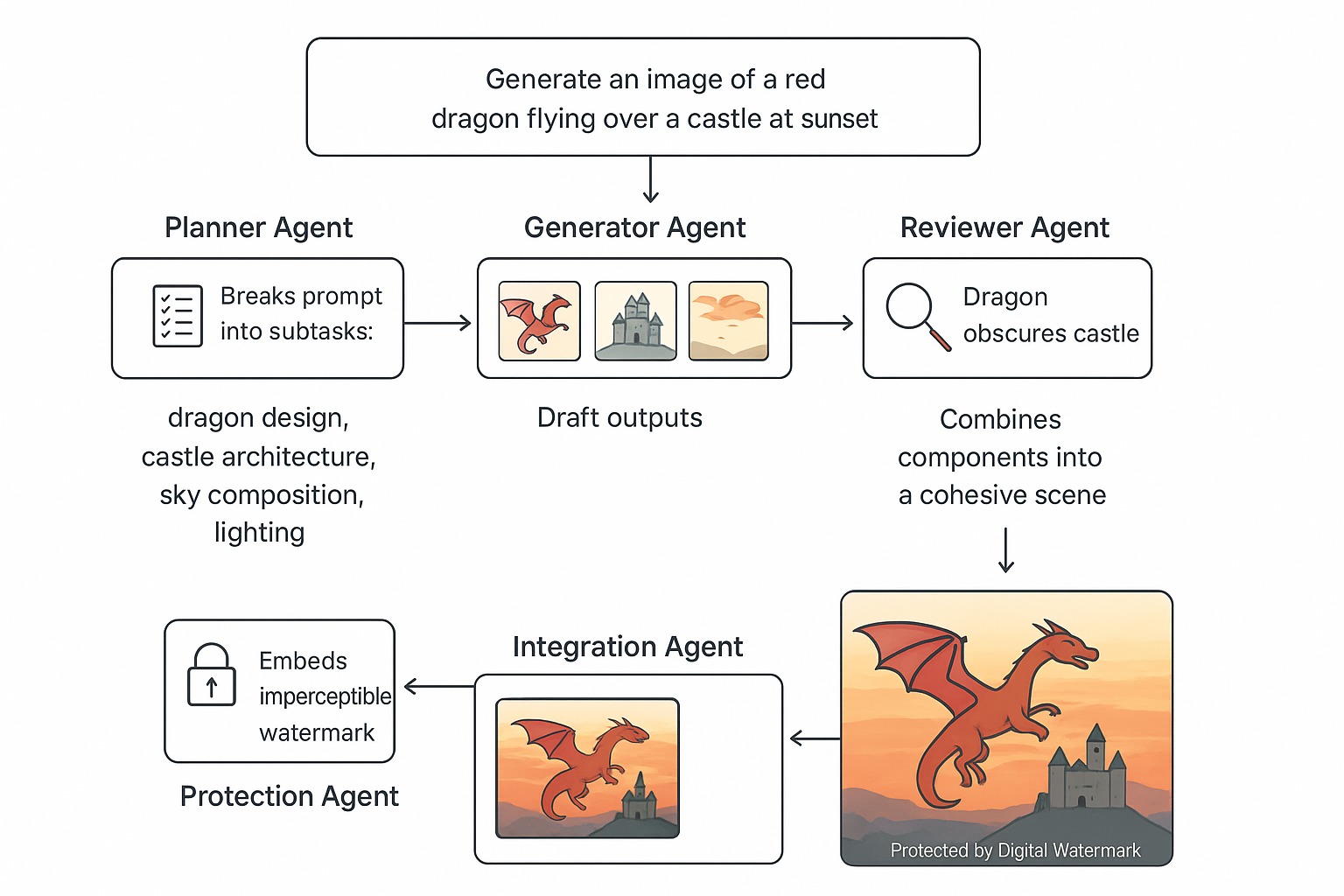}
\caption{Illustrative case study showing the multi-agent pipeline processing the prompt "Red dragon flying over a castle at sunset." The framework decomposes, generates, reviews, integrates, and protects content with embedded watermarking.}
\label{fig:casestudy}
\end{figure}

To demonstrate the framework's creative capabilities, consider the request: "Generate an image of a red dragon flying above a medieval castle during a dramatic sunset." Fig. \ref{fig:casestudy} illustrates the complete workflow.

The Planner decomposes this into subtasks: dragon design (pose, color, scale), castle architecture (style, positioning), sky composition (sunset colors, clouds), and overall layout. The Generator creates initial versions using targeted prompts derived from this decomposition. The Reviewer evaluates outputs using CLIP-based alignment scoring, detecting issues such as the dragon's positioning obscuring castle details. The Integration agent adjusts composition to achieve visual balance while preserving dramatic impact. Finally, the Protection agent embeds an imperceptible watermark and records provenance metadata.

This case demonstrates how task decomposition enables precise control over complex creative requests while maintaining artistic coherence. Users can intervene at any stage to refine specific elements without restarting the entire process.

\subsection{Case Study 2: Copyright Protection in Commercial Applications}

To address practical legal concerns, consider a commercial design agency using AI-generated artwork for client projects. Traditional approaches create vulnerability: clients cannot verify content originality, and the agency lacks proof of creation for intellectual property claims.

Our framework addresses these concerns through integrated protection mechanisms. When generating marketing visuals, the Protection agent embeds unique watermarks tied to the client account and project metadata. This creates verifiable provenance chains essential for copyright protection. If the generated artwork later appears without authorization, watermark detection can establish ownership and creation timestamp.

Recent legal cases highlight this need. In 2023, several high-profile disputes arose over AI-generated artwork used in commercial campaigns without proper attribution or licensing \cite{yu2021artificial}. Our approach provides technical infrastructure to support legal frameworks by ensuring every generated piece contains verifiable ownership information resistant to common transformations (scaling, cropping, format conversion).

The framework also addresses the "derivative work" question in AI-generated content. By maintaining detailed logs of the generation process, including source prompts, intermediate outputs, and refinement steps, our system provides evidence of creative transformation that may be crucial in fair use determinations.

\subsection{Implementation and Feasibility Evidence}

Our prototype demonstrates immediate feasibility using existing components: GPT-4 for planning, Stable Diffusion XL for generation, CLIP for review, standard compositing for integration, and recent diffusion watermarking methods for protection \cite{chen2025robust}.

Table \ref{tab:feasibility} presents feasibility evidence from prior work. Task decomposition approaches like MuLan show 20-25\% CLIPScore improvements over single-step generation \cite{li2024mulan}. Integrated watermarking achieves >90\% recovery rates under JPEG compression, significantly outperforming post-hoc methods at ~70\% \cite{chen2025robust}. Human-in-the-loop studies indicate that reviewer feedback reduces user iterations from 4-5 to 2-3 for satisfactory outputs \cite{wang2022human}.

\begin{table}[!t]
\centering
\caption{Feasibility Evidence from Prior Work}
\label{tab:feasibility}
\begin{tabular}{@{}lcc@{}}
\toprule
\textbf{Evaluation Aspect} & \textbf{Baseline} & \textbf{Improvement} \\
\midrule
Controllability (CLIPScore) & Single-step gen & +20-25\% \\
Watermark recovery (JPEG) & Post-hoc (~70\%) & Integrated (>90\%) \\
User iterations to satisfaction & 4-5 iterations & 2-3 iterations \\
\bottomrule
\end{tabular}
\end{table}

\section{Discussion and Conclusion}

\subsection{Strengths and Applications}

Our multi-agent approach offers significant advantages for practical generative AI deployment. The modular architecture enables targeted optimization while maintaining coherence, crucial for professional workflows. Built-in protection mechanisms ensure content ownership without compromising quality, addressing legal vulnerabilities in commercial applications. The interactive control system provides unprecedented user agency, essential for brand consistency and creative precision.

From a legal and ethical perspective, the framework aligns with emerging requirements for AI transparency and accountability. The detailed provenance logging supports intellectual property claims, while watermarking provides technical infrastructure for copyright enforcement.

\subsection{Limitations and Future Work}

Current limitations include computational overhead from multi-agent orchestration, potentially impacting real-time applications. Watermark robustness against sophisticated adversarial attacks requires ongoing investigation, as advanced attacks targeting watermark removal pose challenges \cite{saberi2024robustness}.

User experience evaluation remains limited to preliminary testing. Large-scale studies across diverse creative domains and user populations are necessary to validate general applicability and identify optimization opportunities.

Future work will extend the framework to video and audio generation, integrate advanced retrieval components for knowledge-grounded creation, and investigate enterprise deployment scalability. We also plan to explore integration with legal frameworks for automated compliance checking and rights management.

\subsection{Conclusion}

This paper presents the first framework combining multi-agent controllable generation with integrated content protection, addressing critical needs in responsible generative AI deployment. Our approach demonstrates how agent-based decomposition can improve user control while built-in watermarking ensures content ownership and provenance tracking.

The preliminary feasibility demonstration shows promising results for both controllability and protection objectives, with practical applications in creative industries and legal contexts. As generative capabilities become increasingly central to information systems, frameworks balancing creativity with protection and control will prove essential for maintaining user trust and legal compliance.

We call upon the research community to explore the intersection of responsible generative AI and practical deployment challenges. Technical solutions must address not only generation quality but also the legal, ethical, and commercial requirements that govern real-world applications.

\end{document}